\documentclass{sig-alternate-05-2015}

\usepackage{balance}

\begin{document}

% Copyright
%\setcopyright{acmcopyright}

\newcommand\acmConference[4][]{%
  \gdef\acmConference@shortname{#1}%
  \gdef\acmConference@name{#2}%
  \gdef\acmConference@date{#3}%
  \gdef\acmConference@venue{#4}%
  \ifx\acmConference@shortname\@empty
    \gdef\acmConference@shortname{#2}%
  \fi}

\setcopyright{rightsretained}
\acmConference{ICTIR' 17 Workshop on Search-Oriented Conversational AI (SCAI' 2017)}{October 1, 2017}{Amsterdam, Netherlands}

%\setcopyright{acmlicensed}
%\setcopyright{rightsretained}
%\setcopyright{usgov}
%\setcopyright{usgovmixed}
%\setcopyright{cagov}
%\setcopyright{cagovmixed}

% DOI
%\doi{10.475/123_4}

% ISBN
\isbn{123-4567-24-567/08/06}

%Conference
%\conferenceinfo{PLDI '13}{June 16--19, 2013, Seattle, WA, USA}

\acmPrice{\$15.00}

%
% --- Author Metadata here ---
\conferenceinfo{SCAI}{'17 Amsterdam, Netherlands}
%\CopyrightYear{2007} % Allows default copyright year (20XX) to be over-ridden - IF NEED BE.
%\crdata{0-12345-67-8/90/01}  % Allows default copyright data (0-89791-88-6/97/05) to be over-ridden - IF NEED BE.
% --- End of Author Metadata ---

%\title{Conversational Information Navigation}
\title{LD-SDS: Towards an Expressive Spoken Dialogue System\\ based on Linked-Data}
%\subtitle{[Extended Abstract]
%\titlenote{A full version of this paper is available as
%\textit{Author's Guide to Preparing ACM SIG Proceedings Using
%\LaTeX$2_\epsilon$\ and BibTeX} at

\numberofauthors{5} %  in this sample file, there are a *total*
\author{
% 1st. author
\alignauthor
Alexandros Papangelis \\
			 \affaddr{Speech Technology Group,}
       \affaddr{Toshiba Research Europe}
       \email{{\small alex.papangelis@crl.toshiba.co.uk}}
% 2nd. author
\alignauthor
Panagiotis Papadakos \\
       \affaddr{Institute of Computer Science, FORTH-ICS, GREECE}
       \email{{\small papadako@ics.forth.gr}}
% 3rd. author
\alignauthor Margarita Kotti\\
			 \affaddr{Speech Technology Group,}
       \affaddr{Toshiba Research Europe}
       \email{{\small margarita.kotti@crl.toshiba.co.uk}}
\and  % use '\and' if you need 'another row' of author names
% 4th. author
\alignauthor Yannis Stylianou\\
			 \affaddr{Speech Technology Group,}
       \affaddr{Toshiba Research Europe,}
			 \affaddr{and Computer Science Department, University of Crete, GREECE}
       \email{{\small yannis.stylianou@crl.toshiba.co.uk}}
% 5th. author
\alignauthor Yannis Tzitzikas\\
       \affaddr{Institute of Computer Science, FORTH-ICS, GREECE,}
       \affaddr{and Computer Science Department, University of Crete, GREECE}
       \email{{\small tzitzik@ics.forth.gr}}
\alignauthor Dimitris Plexousakis\\
       \affaddr{Institute of Computer Science, FORTH-ICS, GREECE,}
       \affaddr{and Computer Science Department, University of Crete, GREECE}
       \email{{\small dp@ics.forth.gr}}
}

\maketitle

\newcommand{\hippalus}[0]{{\tt Hippalus}}

\begin{abstract}
In this work we discuss the related challenges and describe an approach towards the fusion of state-of-the-art technologies from the Spoken Dialogue Systems (SDS) and the Semantic Web and Information Retrieval domains. We envision a dialogue system named LD-SDS that will support advanced, expressive, and engaging user requests, over multiple, complex, rich, and open-domain data sources that will leverage the wealth of the available Linked Data. Specifically, we focus on: a) improving the identification, disambiguation and linking of entities occurring in data sources and user input; b) offering advanced query services for exploiting the semantics of the data, with reasoning and exploratory capabilities; and c) expanding the typical information seeking dialogue model (slot filling) to better reflect real-world conversational search scenarios.

%c) advancing the available SDS technologies to take advantage of the rich, domain independent datasets and the advanced query services in an efficient way; and d) propose an innovative way of communication of the SDS with a live database, instead of a simple ontology or a static knowledge graph. Finally, we describe some preliminary results.

\end{abstract}


%\ccsdesc[500]{Computer systems organization~Embedded systems}


%
%  Use this command to print the description
%
\printccsdesc


\section{Introduction}

Conversational systems is a thriving research area with many commercial applications, such as intelligent personal assistants, e.g.  Microsoft's Cortana, 
Apple's Siri, and Amazon's Echo among others. In addition, Amazon Lex and Watson Conversation have made the functionality of building, testing, and deploying chatbots publicly  available. However, most deployed systems still utilise the rather cumbersome and time-consuming process of hand-crafted rules and finite state machines for dialogue management, paired with domain specific and sometimes static databases.

Moving away from commercial applications, dialogue managers are typically statistical and require a lot of data -- either crowd-sourced or simulated -- to train their internal models, which is a challenging task. Various toolkits are available for developing such systems, e.g.: OpenDial \cite{RapidPrototyping:Stoyanchev2016} is a web-based tool that allows a user to create a slot filling dialogue system automatically, using about fifteen probabilistic rules; Olympus is a complete framework for implementing spoken dialogue systems (SDS) \cite{bohus2007olympus}; PyDial \cite{ultes2017pydial} is a toolkit for developing and training statistical multi-domain SDS based on the slot filling paradigm. Lately, \cite{HybridCodeNetworks:Williams2017} have proposed a  way to reduce the amount of training data by encoding domain-specific knowledge and using developer-provided action templates.

Information-seeking (usually modeled as slot filling, see Section \ref{sec:isd}) is a very common use case for SDS, where a database is inquired for a specific item given a set of hard restrictions \cite{ScalingUpDeep:Cuayahuilty2017, ANetworkBased:Wen2017}. However, databases in such systems are usually small and domain-specific. Efforts to advance such task-based systems include hybrid approaches, for example, in \cite{CombiningChatandTask:Papaioannou2017} slot filling is combined with a chatbot. The aim is to have more natural and engaging systems that switch between task-based and chatbot style dialogue, leading to more satisfied users. Another way to improve the slot filling dialogue experience is via adding memory; for example, in \cite{End-to-endJoint:HakkaniTur2017} limited contextual dialogue memory is used to jointly optimise the signals of user slot filling, intent prediction and system action prediction. Memory is essential, especially for commercial applications, as underlined by \cite{AFrameTrackingModel:Schulz2017}. In their system, the user can refer back to an earlier state in the dialogue, which is essential e.g. when comparing alternatives or researching a complex subject. Recently, efforts are being made towards multi-domain SDS, as well as SDS that work with large knowledge graphs that combine independent domain-specific databases with other knowledge bases (e.g. Freebase) \cite{ConstraintBasedOpenDomain:Aghaebrahimian2016, DialogueManager:Gasic2017, Papangelis:2016, Papangelis:2017}. A general trend, therefore, is to move to conversational open-domain question answering (QA). Large-scale QA systems like IBM's DeepQA rely on multiple sources to form their response: Wikipedia, other knowledge bases, dictionaries, news articles, books, etc. However, the dialogue is usually limited to one turn.

As detailed above, current systems are usually: i) statistical on small domains; or ii) hand-crafted on large/open domains, mostly offering QA, rather than a natural conversation. In this work, we propose an architecture that combines the benefits of SDS and QA, and allows users to retrieve meaningful information by conversing with the system. We achieve this by taking advantage of Linked-Data, Semantic Web, and Information Retrieval technologies. Specifically, our Linked-Data SDS (LD-SDS) can fetch items from available open Linked-Data sources and support advanced exploratory and query services that will help users better understand their own goals and criteria, which are not always explicit and predefined. We here present a prototype of this system, which in the future will have memory, via exploiting a session-based model for exploratory search. Our vision is to enable users to make general queries that lead to more natural conversations.

The rest of the paper is organized as follows: Section \ref{sec:Back} provides the related background regarding the exploratory nature of most information needs and current exploratory approaches, Linked-Data and integration of data from different resources, and the information seeking dialogue models. Section \ref{sec:architecture} discusses the proposed architecture and the prototype we have developed, while Section \ref{sec:challenges} presents the related challenges. Finally, Section \ref{sec:conc} concludes the paper.

\section{Background}
\label{sec:Back}

\subsection{Information Needs \& Exploration}
\label{sec:InformationExploration}

In general, we can classify information needs into two very broad categories: a) {\em precision-oriented}; and b) {\em recall-oriented}. Precision-oriented information needs typically are not time consuming and the goal is to locate one resource and/or its attributes like a telephone number, an address, etc. On the other hand, recall-oriented needs frequently are time expensive,  and the goal is to locate and analyze/compare a set of resources and/or their attributes, or aggregated attributes or interelationships. Recall-oriented tasks aim at {\em decision making} over one or more criteria and have an exploratory nature, like search tasks in the medical, legal, consulting, patent, or academic fields, and consumer-related tasks like car buying, travel planning or even species identification. The following key attributes of exploratory tasks have been identified in \cite{wildemuth2012assigning}: a) they are associated with the goals of learning and/or investigation, b) they are general rather than specific, c) they are open-ended, d) they target multiple items, e) they involve uncertainty, f) they elicit through ill-structured information problems, g) they are dynamic, h) they are lengthy, i) they are multi-faceted, j) they are complex, and finally k) they are accompanied by other information and cognitive behaviors, like sense-making.

According to Marchionini \cite{marchionini2006exploratory} the majority of information needs are {\em recall-oriented}. In the same direction, Broder \cite{broder2002Taxonomy} categorizes queries as navigational (e.g. ``Porsche site"), informational (e.g. ``what is the best mobile phone") and transactional for performing a task (e.g. ``book a hotel"). According to the same author, the queries that are related to recall-oriented needs (i.e. the  informational and transactional queries) correspond to 80\% of queries (50\% for the first and 30\% for the latter query category).
Conversational needs have an exploratory nature and are recall-oriented.

Despite this, current information systems like general purpose web search engines mainly focus on single query precision-oriented needs. Only a small number of prototype information systems provide means for supporting recall-oriented ones (e.g. \cite{papadakos2012exploiting,  papadakos2014hippalus, qarabaqi2014user}). Most of these systems offer their exploratory features (e.g. overviews of available objects, their active attributes/values, counts, etc.) on top of faceted search, an interaction framework based on a multi-dimensional classification of data objects that allows a guided, yet unconstrained way of browsing and exploring the information space through a simple user interface. Faceted search is currently the de facto standard in e-commerce (e.g. eBay, booking.com), and its popularity and adoption is increasing in several other domains \cite{tzitzikas2017faceted}. Features of this framework include: (a) display of current results in multiple categorization schemes either statically or dynamically mined through structured or unstructured data sources \cite{papadakos2012exploiting} (called facets, or dimensions, or attributes, or slots) through the selection of hard constraints, (b) display of facets and values leading to non-empty results only, (c) display of the count information for each value (i.e. the number of results the user will get by selecting that value), and (d) the ability to refine the focus gradually, since it is a session-based interaction paradigm.

Preference-enriched Faceted Search (PFS) \cite{tzitzikas2013interactive} is an extension of faceted search that supports the expression of soft constraints (i.e. preferences) that can impose a ranking over the facets, values and objects of the provided information space. PFS is aligned with the principles of faceted search, allowing users to define explicitly the desired preference structure in a gradual and flexible manner, supporting also set-valued attributes and hierarchically organized values. In brief, PFS adds actions that allow the expression of  various and even conflicting preferences for ranking facets, values, and objects. Such preference actions include best/worst values, relative preferences (e.g. I prefer A to B), around/not around and between actions. PFS can exploit a number of policies for composing preference actions over different facets (e.g. priority, skyline\footnote{Database operator that filters out results, keeping only those objects that are not worse than any other object on all criteria}, etc.). \hippalus\ is a system that implements PFS, as described in Section \ref{sec:architecture}.

\subsection{Linked-Data and Data Integration}
\label{sec:LinkedData}

A big number of datasets (or sources) has been published according to the principles of Linked Data and this number keeps increasing. The ultimate objective of LOD (Linked Open Data) is linking and integration, for enabling discovery and integrated query answering and analysis.
Linked Data refers to a method of publishing structured data, so that they can be interlinked and become more useful through semantic queries, founded  on  HTTP, RDF\footnote{https://www.w3.org/RDF/} and URIs.  As proposed in \cite{bizer2009linked} the major principles of Linked Data are the use of HTTP URIs as names for things, the use of standards for providing useful information and the inclusion of links to other URIs to help the exploration of even more resources.

Therefore, one of the major principles of Linked Data, i.e. interlinking of URIs, favor integration by proposing to the publishers to establish relationships with URIs from other datasets. The linking  of datasets is essentially realized through the existence of common URIs, referring to schema (defined through  RDF Schema\footnote{http://www.w3.org/TR/rdf-schema/} and OWL\footnote{http://www.w3.org/TR/owl2-overview/}) or data elements. SPARQL is a standard query language for retrieving and manipulating RDF data across diverse data sources. The need for better linking and contextualization is also evident from proposals that rate open data, such as the  5-star Open Dat\footnote{http://5stardata.info/en/}, as well as ratings for  vocabulary use \cite{janowicz2014five}. According to \cite{ermilov2016lodstats}, the LOD cloud already contains over 9,000 datasets and billions of RDF triples.

Note that before taking advantage of the available LOD sources there is a preparation phase that generates the semantic layer. During this phase and depending on the exploitation scenario and the properties of those data sources, we have to decide whether we should build a semantic warehouse that will host the gathered triples (e.g. as in \cite{tzitzikas2016unifying,isjwis2016connectivity}) or  a mediated access layer (e.g. as in  \cite{fafalios2013xens}). A dedicated warehouse has the additional benefit of gaining reliability and efficiency -- for example the SPARQL endpoint provided by DBpedia is not very stable or fast.

This semantic data layer can be used for both analyzing what the user says as well as for feeding the responses of the SDS. Specifically, it can be exploited for identifying named entities, for getting their descriptions or related entities through the semantic data layer either from the user's response or other unstructured data sources. This is sometimes referred as semantic enrichment \cite{fafalios2015exploiting}. In addition, the semantic layer can be used for producing summaries by estimating the more important elements or associations \cite{fafalios2014enriching}.

\subsection{Information-Seeking Dialogue Models}
\label{sec:isd}
Dialogue models for information-seeking applications have been traditionally modeled with the slot filling paradigm \cite{Surdeanu:2014}. According to that, the goal of the dialogue is to extract enough information from the user in order to form a database query that yields acceptable results, e.g. according to the number of items returned or some other metric. The database's attributes, therefore, are called slots, and the process of forming the query through dialogue is called slot filling. According to \cite{Surdeanu:2014}, each slot $s$ can take one value $v_s$ from a given set of values: $v_s \in V_s, s \in S$, where $V_s$ may represent a set of strings, events, numerical values, general entities, etc., and $S$ is the set of slots. An \emph{ontology} describes each slot and its values, as well as other attributes of the database's schema. 

Without loss of generality, we can assume that $V_s, \forall s \in S$ is a set of symbols, somehow extracted by Spoken Language Understanding (SLU), representing something meaningful for each slot. Usually, it is assumed that the goal of the user can be expressed as a tuple of slot-value pairs: $<s_0=v_0, ..., s_k=v_k>$, a representation that may be somewhat restrictive if we hope to train SDS for realistic applications. Indeed, in recent work, especially for dialogue management, a simple version of slot filling is typically adopted to model human - machine dialogues, e.g. \cite{DialogueManager:Gasic2017,Papangelis:2017,ANetworkBased:Wen2017,HybridCodeNetworks:Williams2017}.

Such approaches to model information-seeking dialogues have been largely driven by the trend to view dialogue as an optimisation problem and address it with (Partially Observable) Markov Decision Processes (POMDP). Specifically, a POMDP Dialogue Manager (DM) typically receives an n-best list of language understanding hypotheses, which are used to update the belief state (reflecting an estimate of the user's goals). Using Reinforcement Learning (RL), the system selects a response that maximises the long-term return of the system. This response is typically selected from an abstract action space and has to be converted to text through language generation. %Concretely, a POMDP is defined as a tuple $<Z,A,T,O,\Omega, R, \gamma>$, where $Z$ is the state space, $A$ is the action space, $T:Z\times A \rightarrow Z$ is the transition function, $O:Z \times A \rightarrow \Omega$ is the observation function, $\Omega$ is a set of observations, $R:Z\times A \rightarrow \Re$ is the reward function and $\gamma \in [0,1]$ is a discount factor of the expected cumulative rewards $J=E[\sum_t{\gamma^tR(z_t,a_t)}]$. A policy $\pi:Z \rightarrow A$ dictates which action to take from each state. An optimal policy $\pi^\star$ selects an action that maximises the expected returns of the POMDP, $J$. Learning in RL consists exactly of finding such optimal policies; however, due to state-action space dimensionality, approximation methods or other ways to improve computational complexity need to be used for practical applications.

We here extend the information-seeking model used in most (PO)MDP and Deep Learning approaches to dialogue and connect to live semantic knowledge bases, aiming to create a more realistic paradigm that will drive our research forward. In the following section we describe LD-SDS and then proceed to outline some of the challenges we are facing.

\section{LD-SDS Architecture \& Prototype}
\label{sec:architecture}
Moving towards a more realistic information-seeking dialogue paradigm, it is important to allow users the freedom they need to express their complex intents. These intents do not always express hard restrictions (constraints) but often express preferences that users may or may not be willing to relax as the dialogue progresses. Such preferences may refer to the importance of attributes over other attributes (e.g. \emph{location} is much more important than \emph{has-free-wifi} when searching for accommodation), or may refer to preferred values of a given attribute (e.g. prefer \emph{central} over \emph{northern} locations but \emph{northern} may still be okay under certain circumstances), etc.

We therefore implement a number of operators $Op = \{<, \leq, >, \geq, =, \neq, $\emph{around, not around, between, not between, prefer}$\}$, for two types of constraints: hard and soft. Hard constraints restrict the information space, while soft constraints express preferences and impose a ranking on the available options. Of course, not all constraints are applicable to all types of slots (e.g. if a slot's values are not ordinal). Instead of forcing each slot to have one value, therefore, we allow constraints of the form $s$ $op$ $v_s$, where $op$ is an operator in $Op$ except \emph{between} and \emph{prefer}, for which we have: $s$ $op^\prime$ ($v^1_s$, $v^2_s$). The \emph{prefer} operator is defined for slots in two ways: \emph{prefer X over Y}, which means that slot X is more important than slot Y (this affects the ranking of the results, among other things); and \emph{prefer X and Y}, which means that both slots X and Y are preferred over the rest of the slots available. In a similar manner \emph{prefer} can be defined for a specific slot's values. Besides \emph{not} which is explicitly defined as a different operator to make belief tracking easier, other logical operators such as \emph{and, or} are handled by the SLU and mapped into a set of constraints using the operators described above.

\begin{figure}
\centering
	\includegraphics[scale=.4]{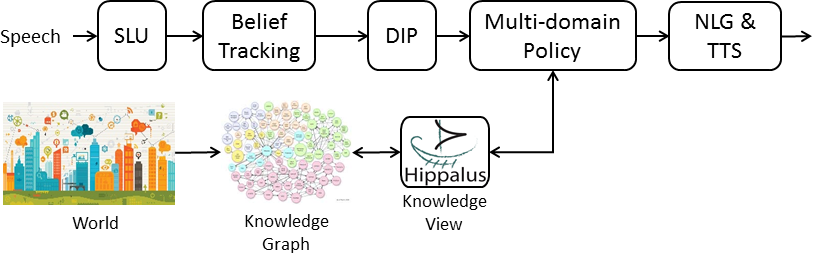}
	\caption{The proposed LD-SDS architecture. TTS refers to Text To Speech synthesis.}
	\label{fig:arch}
	\vspace{-2mm}
\end{figure}

Another aspect of (the typical version of) the slot filling paradigm that we extend is the nature of each slot's values. Specifically, we allow values to be defined in a hierarchical manner (e.g. \emph{location} with regions, sub-regions, neighbourhoods, etc), and we allow slots to take multiple values, from a given set (e.g. \emph{hotel amenities}). Slots with hierarchical or multiple values directly impact SLU and dialogue state / belief tracking (DST / BT). Formally, a slot with hierarchical values can be defined as $s^h = v^h, v^h \in V^H$, where $V^H$ are the nodes of a hierarchy (e.g. a tree). We treat such slots as regular slots (i.e. that take a single value) but handle belief tracking in a way that accounts for the value hierarchy. Multi-valued slots can be defined as $s^m = v^m, v^m \subseteq V^m$, where $V^m$ is the set of acceptable values for slot $s^m$.

To process the complex intent of the user, we connect our SDS with \hippalus, an exploratory search system that materializes the Preference-enriched Faceted Search \cite{tzitzikas2013interactive} over semantic views gathered from different data sources through SPARQL queries. \hippalus\ is a publicly accessible web system\footnote{http://www.ics.forth.gr/isl/Hippalus/} that supports the previously defined hard and soft restriction actions that allow the user to order facets, values, and objects. All the above functionality is offered in an efficient way, by using the algorithms described in \cite{tzitzikas2013interactive}. The information base that feeds \hippalus\ is represented in RDF/S (using a schema adequate for representing objects described according to dimensions with hierarchically organized and set-valued attributes). For loading and querying such information, \hippalus\ supports a number of triple stores. The performed actions are internally translated to statements in the preference language described in \cite{tzitzikas2013interactive}, and are then sent to the server  through HTTP requests. The server analyzes them,  using the language's parser, and checks their  validity. If valid, they are passed  to the appropriate preference algorithm. Finally, the  respective preference bucket\footnote{A preference bucket holds incomparable objects regarding the given soft-constraints} order  is computed and the ranked list of objects according to preference is sent to the user's browser. The answer is also enriched with a number of metrics computed over the response (preference score, pair-wise wins of buckets' objects regarding preference,  etc.) that can be exploited by the Belief Tracking and Natural Language Generation components (e.g. for identifying the most important slots and breaking ties through user input). Figure \ref{fig:arch} shows the overall architecture of the system.

\subsection*{Example Interaction}

\begin{figure}
\centering
	\includegraphics[scale=.190]{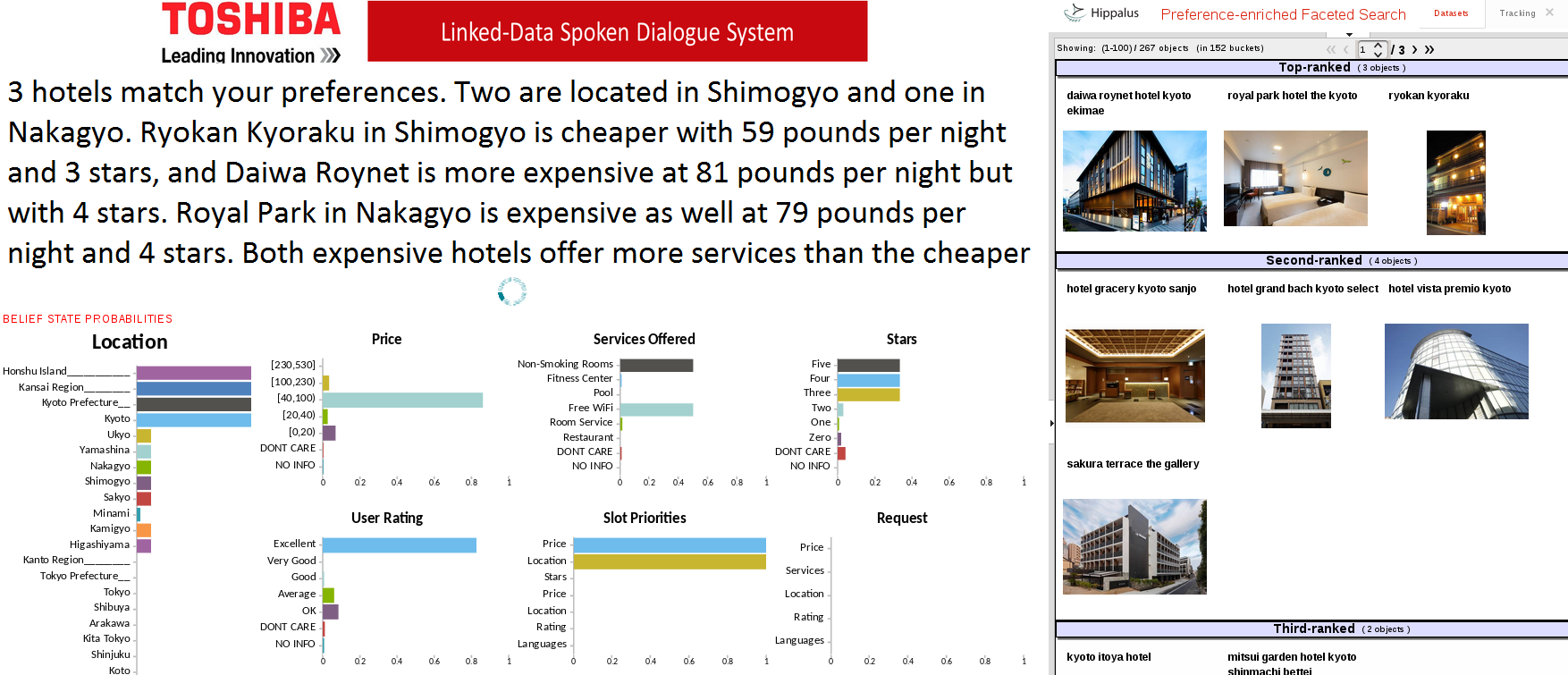}
	\caption{Screenshot of LD-SDS. Parts of the belief space on the left, Hippalus output on the right.}
	\label{fig:screen}
	\vspace{-2mm}
\end{figure}

Figure \ref{fig:screen} shows a screen shot of the working LD-SDS prototype, on a knowledge base about Japanese Hotels (382 hotels across Japan, 69 system actions, and 28 slots some of which are hierarchical and some are multi-valued). To showcase our conversational information navigation paradigm, we walk through a real interaction with our prototype LD-SDS in Table \ref{tab:example}. Belief tracking is performed as described in the following section; parallel to this we also process the recognised dialogue acts to identify constraints or preferences. In the example we also note very briefly the gist of the response from Hippalus. 

\begin{table}[!ht]
\begin{center}
	\begin{tabular}{|p{\linewidth}|}
		\hline
		{\bf System}: Hello, welcome to Toshiba's Japanese Hotels spoken dialogue system. How may I help you? \\
		\hline
		{\bf User}: I'm looking for a hotel in Kyoto but not in Minami where they offer free Wi-Fi and have non smoking rooms. (ASR confidence: 0.75154209)  \emph{Minami is a region in Kyoto.} \\
		\hline
		{\small {\bf Constraints}: [ type=hotel, location=kyoto, location$\neq$minami, amenities=\{free-wifi, non-smoking-rooms\} ]} \\
		{\small {\bf Preferences}: [ ]} \\
		{\small {\bf Hippalus}: [ 267 items in 2 buckets; \{60, 207\} items in buckets ]} \\
		\hline
		\hline
		{\bf S}: What price range are you looking for? \\
		\hline
		{\bf U}: I want something around 70 pounds and with more than two stars. (0.7065863) \\
		\hline
		{\small C: [ pricerange $\sim$ 70, stars > 2 ]} \\
		{\small P: [ ]} \\
		{\small H: [ 188 items in 45 buckets; \{25, 103, 13, ...\} items in buckets ]} \\
		\hline
		\hline
		{\bf S}: What kind of user rating would you prefer the hotel to have? \\
		\hline
		{\bf U}: I'd like excellent ratings. (0.92533112)\\
		\hline
		{\small C: [ ]} \\
		{\small P: [ ratings=excellent : best ]} \\
		{\small H: [ 188 items in 27 buckets; \{9, 11, 8, ...\} items in buckets; identified slots that can affect ranking and reduce items in first bucket: \{stars, location, price\} ]} \\
		\hline
		\hline
		{\bf S}: Which of the following criteria are important for you? stars, price, or location? \\
		\hline
		{\bf U}: I prefer location and price. (0.95948964)\\
		\hline
		{\small C: [ ]} \\
		{\small P: [ location $>>$ all, price $>>$ all ]} \\
		{\small H: [ 188 items in 152 buckets; 3 items in first bucket ]} \\
		\hline
		\hline
		{\bf S}: 3 hotels match your preferences. Two are located in Shimogyo and one in Nakagyo. Ryokan Kyoraku in Shimogyo is cheaper with 59 pounds per night and 3 stars, and Daiwa Roynet is more expensive at 81 pounds per night but with 4 stars. Royal Park in Nakagyo is expensive as well at 79 pounds per night and 4 stars. Both expensive hotels offer more services than the cheaper one. \\
		\hline
		{\bf U}: Thank you, goodbye. (0.97125274)\\
		\hline
	\end{tabular}
\end{center}
\label{tab:example}
\caption{Example interaction between a human user and our LD-SDS prototype. In the interest of space, the notes under each dialogue turn briefly show items that correspond to new information. The belief state is updated accordingly (see Section \ref{sec:challenges}). ASR: Automatic Speech Recognition.}
\vspace{-2mm}
\end{table}

\section{Challenges}
\label{sec:challenges}
Conversational search is much more complex than traditional (e.g. command line) search, as the algorithms need to take into account what has happened so far in the interaction, need to include common knowledge, and knowledge about the specific user. A SDS should be able to alleviate some of this burden, and this is one of the aims of our work. Moreover, the increased expressiveness necessary for information navigation is more natural in spoken rather than written language, and this may be one way to get SDS more widely adopted and perhaps on par with command line search or apps. In this context, there are a number of challenges that need to be taken into account, all of which fall in our current and future work plans:

{\bf Language Understanding.} In order to process input that reflects complex user intents (and may have different meanings for different users or context, etc.), besides the usual challenges, SLU needs to have access to multiple kinds of knowledge, across many domains. As the system is connected to live knowledge bases, it is important for SLU to be able to adapt over time, as well as handle out of domain input gracefully. 

{\bf Belief Tracking.} Translating the identified user intentions from SLU into a belief state is not trivial, even for slot filling models with one or two operators (e.g. $=, \neq$). As an initial approach to belief tracking under this expanded paradigm, we follow the simple principles outlined on Table \ref{tab:btrules} in conjunction with an existing belief tracker. While this is straightforward for regular slots, when we have slots with hierarchical values or multi-valued slots, we need a different kind of belief update. Specifically, for hierarchical slots we need to recursively perform the belief update, while still following the basic principles of Table \ref{tab:btrules}. As the constraints become more complex, traversing the hierarchy of values becomes non-trivial. In our prototype, we traverse the hierarchy once for each constraint (relevant to a specific hierarchical slot) and then combine the updates into a single belief update by taking the average for each value. When updating multi-valued slots, we simply divide the probability mass across each value that was mentioned (and not negated), although this may not be optimal.
\begin{table}[!h]
\begin{center}
	\begin{tabular}{|p{0.3\linewidth}|p{0.6\linewidth}|}
		\hline
		$< x, \leq x$ & bias update towards values greater / equal to or greater than $x$ \\
		\hline
		$> x, \geq x$ & bias update towards values less / equal to or less than $x$ \\
		\hline
		$= x, \neq x$ & bias update towards values equal or not equal to $x$ \\
		\hline
		around $x$, not around $x$ & bias update towards values around or not around $x$ - e.g. within one std. dev. \\
		\hline
		between $[x,y]$, not between $[x,y]$ & bias update towards values between or outside of $[x,y]$ \\
		\hline
	\end{tabular}
\end{center}
\caption{Basic principles for our initial belief tracker that is based on the tracker provided by \cite{ultes2017pydial}.}
\vspace{-2mm}
\label{tab:btrules}
\end{table}

{\bf Dialogue Policy.} Robust, scalable and multi-domain policy models are necessary, that take into account what has happened in the dialogue (e.g. belief state), what has happened in similar dialogues with other users (global dialogue history), and also what is currently happening in the real world (e.g. response from knowledge bases). As an initial approach, we plan to apply our Deep Q-Network (DQN) trained multi-domain policy network \cite{Papangelis:2017} on this extended paradigm. The core idea is that the policy model is trained on domain-independent features extracted for each slot at run-time and can thus operate even if the slots (or their values) change dynamically. Such features for example include slot entropy (w.r.t. its values), number of database matches if a slot has a specific value, distribution of values over the database, etc. A policy operating in such a feature space, therefore, is independent of domain-specific slots and values.

{\bf Language Generation.} As also noted by other researchers (e.g. \cite{Asri:2017} when analyzing human to human data), a very important capability of the LD-SDS is to be able to summarize the current state of the results and compare various items on various aspects. Other challenges include the ability to describe items information which may reside in multiple live semantic knowledge bases, whose schemas may change over time. Natural Language Generation (NLG) in LD-SDS is currently done via templates that allow us to compare up to 3 items, on aspects mentioned by the user in the constraints and preferences. Our next step is to collect data and train models for NLG, following works such as \cite{Gkatzia:2016,Lebret:2016,Press:2017,Wen:2016}, and enrich \hippalus ' answer with relevant metrics.

{\bf Semantic Layer.}
The integration of available information is a hot and important topic. Although a big number of datasets has been published according to the principles of Linked Data, there is no evidence regarding the connectivity of the current LOD cloud or  its quality. Aspects to consider include the complementarity, the discovery and selection of datasets, the integration, novelty and provenance of information, the evolution of datasets, the scalability and the efficiency of the approaches. Creating appropriate semantic views based on the user input, that offer advanced query and exploratory services by exploiting entity mining, disambiguation methods, available unstructured data sources and the history and profile of the user, is a challenging task.

{\bf Evaluation.} As there is no clear way to define task success in this setting, traditional metrics like precision and recall need to be modified and appropriate baselines have to be carefully selected. General dialogue quality estimators may still be used to capture the conversational aspect of the interaction, but we need metrics to also measure the quality of information retrieved by the system (such as metrics for interactive or session-based information retrieval \cite{tzitzikas2017faceted}).

\section{Conclusions}
\label{sec:conc}
%We have presented our working prototype LD-SDS which operates on an expanded dialogue paradigm (compared to paradigms that most research groups publish in the state of the art) and connects to live semantic knowledge bases. LD-SDS therefore allows users to explore the information space through conversation and make informed decisions, rather than being fed a suitable answer by the system. We are actively working on training statistical models (deep learning and others) the various parts of the SDS outlined in section \ref{sec:challenges}, starting with data collection and designing and training an appropriate simulated user. 

In this paper we have motivated the need for more expressive SDS that can leverage the wealth of Linked Data (semantic knowledge bases in general) and state-of-the-art exploratory search services.  Towards this direction we have presented our prototype LD-SDS which operates on an expanded dialogue paradigm (compared to paradigms in the literature) and connects to live semantic knowledge bases. Specifically, it adopts and exploits a more expressive data model (multi-valued attributes, hierarchically organized values), supports more complex questions/commands (less than, around), supports a session-based interaction model appropriate for (gradual) decision making that supports preferences (i.e. soft constraints for ranking the available options) that exploits the expressiveness of the data model (preference inheritance is supported, scope-based resolution of conflicts), takes initiatives (based on the focus and the distribution of values) for deciding what ``clarification questions" it should ask the user. Last, we have demonstrated the feasibility of this approach for hotel selection.

There are several directions that are worth further research.  One is to investigate how to exploit global scale semantic indexes, like those proposed in \cite{isjwis2016connectivity}. Another is training statistical models (deep learning and others) for the various parts of the SDS outlined in Section \ref{sec:challenges}; we are currently working on this, starting with data collection and designing and training an appropriate simulated user.

%\section{Acknowledgments}

\balance % GM June 2007
\small
\bibliographystyle{abbrv}
\bibliography{References}

% That's all folks!
\end{document}